\documentclass[journal,10pt]{IEEEtran}

\usepackage{graphicx}
\usepackage{psfrag}
\usepackage{subfigure}
\usepackage{url}
\usepackage{stfloats}
\usepackage{amsfonts,amssymb,amsmath,amsbsy,bm,paralist,theorem,ifthen,color}
\usepackage{array}
\usepackage{calc}
\usepackage{subfig}
\usepackage{longtable}
\usepackage{multirow}
\usepackage{algorithmic,algorithm}
\usepackage{graphics,booktabs,epsfig,enumerate}
\usepackage{afterpage}
\usepackage{float}
\usepackage[switch]{lineno}
\usepackage{cite}
\usepackage{pgfplots} 

\definecolor{orange}{RGB}{255, 165, 0}

\usepackage{cancel}
\definecolor{orange}{RGB}{255,107,0}

\pdfoutput=1

\begin{document}
%
%
%

\title{Beam-space Multiplexing: Practice, Theory, and Trends--\LARGE{\emph{From 4G TD-LTE, 5G, to 6G and Beyond}}} 

\author{Shanzhi Chen, \IEEEmembership{Fellow, IEEE},  Shaohui Sun, Guixian Xu, Xin Su, and Yuemin Cai
	\thanks{This work was supported in part by the National Science Fund for Distinguished Young
Scholars (Grant No. 61425012) and the National Science and Technology Major Project (Grant No. 2016ZX03001017).}
	\thanks{S. Chen, S. Sun, X. Su and Y. Cai are with the State Key Laboratory of Wireless Mobile Communications, China Academy of Telecommunications Technology (CATT), Beijing 100191, China (e-mail: chensz@cict.com, \{sunshaohui,suxi\}@catt.cn, caiyuemin@datangmobile.cn).}
	\thanks{G. Xu is with Department of Engineering, Aarhus University, Aarhus, 8000, Denmark (e-mail: guixian@eng.au.dk). }
}

\maketitle

\begin{abstract}
In this article, {the new term, namely beam-space multiplexing, is proposed for the former multi-layer beamforming for 4G TD-LTE in 3GPP releases.}
We provide a systematic overview of beam-space multiplexing from engineering and theoretical perspectives. Firstly, we clarify the fundamental theory of beam-space multiplexing. Specifically, {we provide a comprehensive comparison with the antenna-space multiplexing in terms of theoretical analysis, channel state information acquisition, and engineering implementation constraints.} Then, we summarize the key technologies and 3GPP standardization of beam-space multiplexing in 4G TD-LTE and 5G new radio (NR) in terms of multi-layer beamforming and massive beamforming, respectively. We also provide  system-level performance evaluation of {beam-space multiplexing schemes} and field results from current commercial TD-LTE networks and field trial of 5G. The practical deployments of 4G TD-LTE and 5G cellular networks demonstrate the superiority of beam-space multiplexing { within the limitations of implementation complexity and practical deployment scenarios. } Finally, the future trends of beam-space multiplexing in 6G and beyond are discussed, including massive beamforming for extremely large-scale MIMO (XL-MIMO), low earth orbit (LEO) satellites communication, data-driven intelligent massive beamforming, and multi-target spatial signal processing, i.e., joint communication and sensing, positioning, etc.
\end{abstract}
\IEEEpeerreviewmaketitle

\vspace{3cm}

\section{Introduction}
 In the past twenty years, physical layer technologies such as multiple access waveform, modulation, coding as well as time and frequency domains multiplexing have been flourishing over the evolution of cellular systems. However, in 2G and 3G/B3G, time-frequency domain technology had been thoroughly explored to increase the system capacity. Instead, multiple-input multiple-output (MIMO), also known as the multi-antenna technology, has great potential to increase spectral efficiency (SE) and system capacity by spatial multiplexing, which has been regarded as the core technology in 4G Long Term Evolution (LTE) systems and 5G new radio (NR). Historically, multi-antenna technology has evolved from passive to active, from two-dimensional (2D) to three-dimensional (3D), from few antennas to large-scale array (massive MIMO) \cite{chen2015comprehensive}.  The performances benefited by MIMO in cellular systems are summarized as follows: 
 \begin{itemize}
 	\item Antenna gain: Array gain from multi-antenna can prolong battery life, extend cellular range, and provide higher throughput.
 	\item 	Diversity gain: Spatial diversity from multi-antenna can improve reliability and robust operation of services.
 	\item Multiplexing gain: Multi-stream transmission can provide higher data rates.
 	\item Interference suppression: Transmit and/or receive (Tx/Rx) interference rejection combination (IRC) can increase the signal-to-interference-plus-noise ratio (SINR) and improve link reliability and robustness.
 \end{itemize}

Spatial multiplexing in MIMO systems provides spatial multiplexing gain, which increases system capacity by subdividing an outgoing signal stream into multiple data pipes yet suffers from power efficiency. Beamforming provides significant array gains, thereby providing an increased signal-to-noise ratio (SNR), also known as power gain. To achieve array gain and multiplexing gain simultaneously, beamforming-based spatial multiplexing or multi-layer beamforming was thus proposed, which serves as a milestone in the development of multi-antenna technology and plays a crucial role in the Third Generation Partnership Project (3GPP) LTE standardization of the enhancement of multi-antenna schemes.

In this article, multi-layer beamforming is termed beam-space multiplexing. It was firstly proposed by the authors’ team of China Academy of Telecommunications Technology (CATT)/Datang in LTE Release 9 (Rel-9) to support single-user/multi-user dual-layer beamforming \cite{R1-091513}. The following Rel-10 can support up to eight layers of beam-space multiplexing that further enhanced the multiplexing gain \cite{TR36912, ChenADPT}. Beam-space multiplexing for coordinate multiple point (CoMP) transmission was then standardized in Rel-11 for multi-cell systems. The target is not only for the mitigation of intra-cell interference but also for that of the inter-cell interference.  Afterwards, the enhancements of downlink four-antenna transmission technology and improvements of CoMP technology were further optimized in Rel-12. Besides that, 3D (including both the horizontal and the vertical dimensions) beam-space multiplexing was also studied \cite{ChenADPT}.

Since 2011, eight-antenna base station (BS) with beam-space multiplexing has been widely deployed in commercial time-division LTE (TD-LTE) 4G cellular networks. However, LTE BSs with frequency-division duplex (FDD) are mainly equipped with 2-antennas and adopted with antenna-space multiplexing. FDD BSs have seldom four antennas. As reported by the Global TD-LTE Initiative (GTI) white paper, TD-LTE cellular networks have 50\% more area coverage than that of LTE FDD in typical deployments, which means that 33\% of BS installation cost can be saved \cite{GTI}. Currently, the total number of 4G commercial BS’s global installation is about 6.31 million. Among them, the number of TD-LTE BSs is 2.89 million, including 1.5 million eight-antenna BSs. There are 4 billion 4G subscribers globally, of which TD-LTE has 1.8 billion subscribers.

In 5G networks, beam, formed by a BS with large-scale antenna array, is much narrower. It provides extremely high directional selectivity and array gain \cite{Tom}. Furthermore, BS operates at mmWave frequencies can pack more antennas within a small size of an antenna array. Built with high spatial resolution, an antenna array can considerably increase the data rate while reducing power consumption. However, the acquisition of channel state information (CSI) becomes an essential issue. In FDD systems, CSI can be obtained through CSI estimation and feedback. As the number of BS antennas is increased, the size of reference signal symbols and feedback cost overheads are unfeasible high. Alternatively, channel reciprocity in time-division duplex (TDD) systems can be used to conquer such problems \cite{ChenADPT}. Thus, TDD systems with beam-space multiplexing in terms of massive beamforming will be the mainstay in development of multiple antenna technology for the upcoming 5G \cite{Chen2014}. Moreover, beam-domain signal processing, especially beam-space multiplexing will play an important role in 6G and beyond, which is viewed as an intelligent network enabled by machine learning (ML).   In this article, we outline several potential research directions and challenges: massive beamforming for extremely large-scale MIMO (XL-MIMO), low earth orbit (LEO) satellite communication, and data-driven intelligent massive MIMO. Besides, multi-target spatial signal processing for joint communication and sensing with massive MIMO needs further {investigations}.

Inspired by the successful engineering practices of beam-space multiplexing, this article intends to provide a comprehensive overview from both engineering and theoretical perspectives. This article will systematically summarize of beam-space multiplexing, from the theoretical analysis to applications in the scope of 3GPP 4G TD-LTE and 5G cellular systems. Furthermore, the emerging massive beamforming in 5G and further research trends in 6G and beyond are envisioned.

{{\textbf{Organization of this article:}}} In the next section, we first introduce fundamental principles of beam-space multiplexing and analyze engineering feasibility related to antenna-space multiplexing. {The standardization, system-level simulations, and field-test results of transmission schemes are summarized with focusing on beam-space multiplexing in 4G TD-LTE and 5G.} Lastly, future research directions related to massive beamforming in 6G and beyond are discussed.


\begin{figure}[!hbt]
	\centering
	\includegraphics[width=0.49\textwidth]{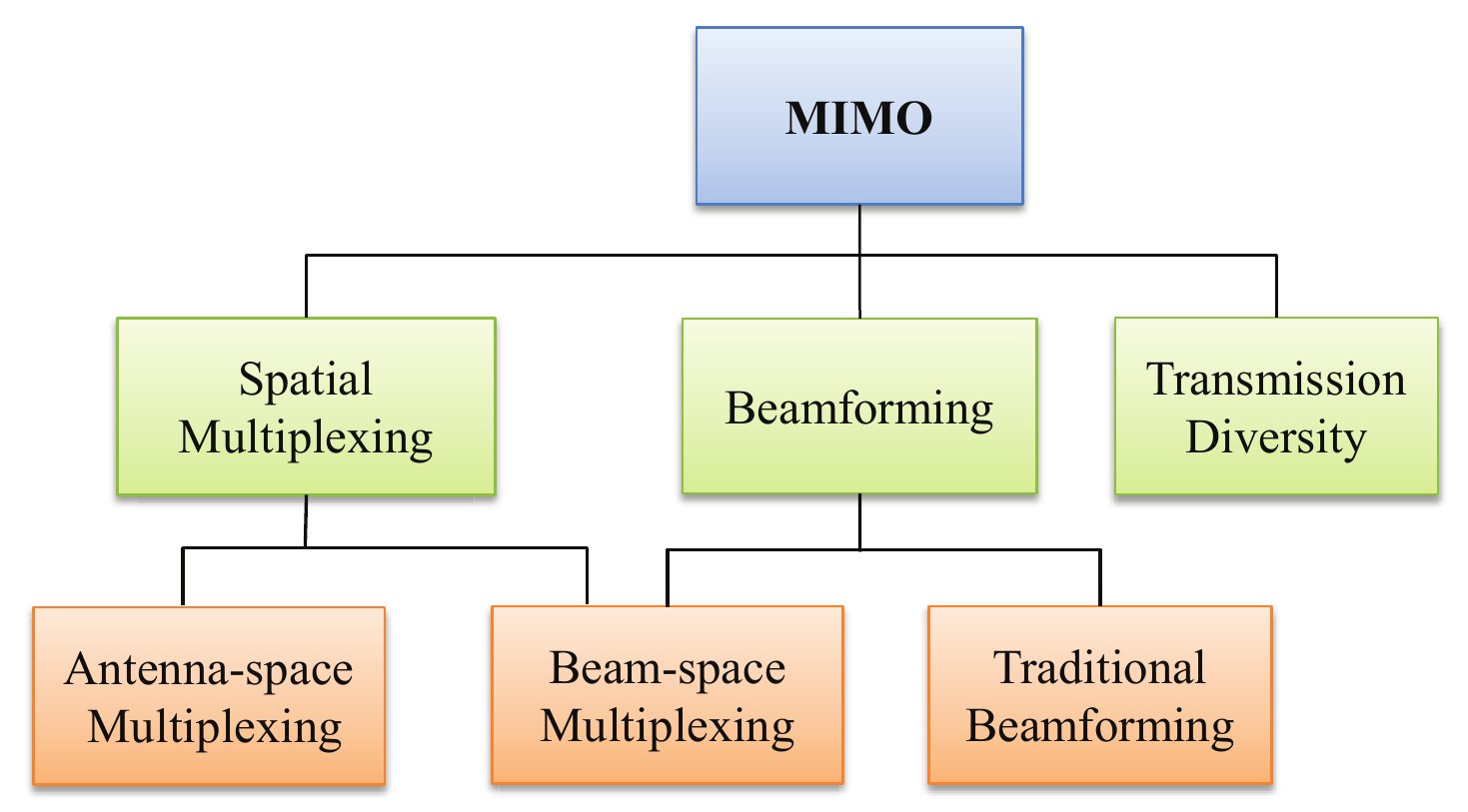}
	\caption{Classification of MIMO technology}
	\vspace{-0.2cm}
\end{figure}\label{Fig:1}

\vspace{0.8cm}
\section{Beam-space Multiplexing}
From the engineering point of view, the authors’ team of CATT/Datang was the first proposer and the major contributor of multi-layer beamforming scheme for 4G TD-LTE standard in 3GPP. It is a breakthrough and key technology of 4G TD-LTE, which has been recognized as a sufficient method to improve SNR and system capacity, especially for cell-edge users. Moreover, it has far-reaching impacts on 5G with practical significance. Note that, this term is related to multiple layers’ data transmission from the application point of view instead of a technical concept. Therefore, we use beam-space multiplexing terminology to signify multi-layer beamforming in this article.

As illustrated in Fig. 1, beam-space multiplexing sufficiently utilizes the advantages of both beamforming and spatial multiplexing. With beamforming techniques, the beam patterns of transmit and receive antenna array can be steered in number of certain desired directions, whereas undesired directions (e.g., directions of significant interference) can be suppressed (‘nulled’). Moreover, spatial multiplexing based on beamformed spatial channel enables multiple activated data pipes, which enhances data rate. 

\vspace{-0.3cm}
\subsection{\textbf {Theoretical Analysis}}
Firstly, we compare the differences between the operational mechanism, signal property, and performance gain concerning various transmission schemes. As highlighted in Table 1, antenna-space multiplexing provides high multiplexing gain that increases the system throughput by transmitting multiple data layers concurrently over multiple parallel subchannels, where each channel can be independent to each other through large-spaced (i.e., $4\!\sim\!10\lambda$)  antennas at the BS. Then, the corresponding spatial signal processing in antenna domain generally creates discretionary spurious signal through precoding. The traditional beamforming is a classical single-beam signal processing technique, where multiple short-spaced (i.e., $\lambda/2$) antenna elements are adaptively phased to form a dedicated beam pattern of direction. Beamforming can be used at BS and user side to improve SNR with significant array gain. Beam-space multiplexing combines the traditional beamforming and spatial multiplexing by exploiting multipath independent propagation capability. Beam-space multiplexing is expected to transmit multiple dedicated directional signals with high array gain. It could obtain high multiplexing gain and power efficiency simultaneously and expand the coverage with high data rate.  {In principle, it would suffer from the power leakage and the inter-user inference if the 
direction of selected beams from the fixed codebook were not exactly point to the user direction and could not eliminate interference to other co-scheduled users. However, if full CSI is available at network side, e.g., in TDD system, such issue can be alleviated with more accurate non-codebook-based beamforming. 
 }

{Then, we first review the multiplexing capacity of deterministic MIMO channels with antenna-space multiplexing.  For antenna-space multiplexing, the capacity for single user with time-invariant channel  is $\sum_{i =1}^{n_{\rm min}} \log_2\left(1+ \frac{p_i^* \mu^2_i}{N_0}\right)$ bits/s/Hz, where ${n_{\rm min}}:= \min(N_t, N_r)$ and $\mu_i$ is the $i$-th singular value (descending order) of channel $\mathbf H$ that represents the spatial channel  with $N_t$  transmit antennas and $N_r$ receive antennas; $N_0$  denotes the noise power at the receiver and $p^*_1, \dots, p_k^*$ are the allocated power  for spatial parallel sub-channels determined by water-filling \cite{Tse}. At high SNR, equal power allocation policy is the asymptotically optimal, the capacity can be derived by
\begin{align}
\mathcal C_{\rm ant-space} & \approx \sum_{i=1}^{k} \log_2 \left(1 + \frac{P_{\rm total}\mu^2_i}{kN_0} \right) \\ \notag
& \approx k \log_2 (\texttt{SNR})+ \sum_{i=1}^{k} \log_2 \left(\frac{\mu^2_i}{k}\right) {\text{bits/s/Hz.}}
\end{align}where $\texttt{SNR} = P_{\rm total}/N_0$ and $k$ also denotes the number of spatial degrees of freedom offered by multiple antennas.}

{With beam-space multiplexing, the received signal vector at the receiver can be expressed as $\mathbf y = \mathbf {HVs+ n}$, where $\mathbf V \in \mathbb C^{N_t \times r}$ 
is the optimal beamforming matrix with $r$ beams such that the spatial multiplexing shifts to beam domain from antenna domain, and $\mathbf n$ is the additive noise at the receiver. When $r = 1$, it is traditional beamforming. Then, the capacity is $\mathcal C_{\rm beam} = \log_2 (1+\texttt{SNR}  ~\mu^2_1)$ such that high power efficiency is gained. 
Assuming the effective channel $\mathbf {HV}$ has $r$ non-zero singular values, and $r \leq k$. Similarly, the achievable capacity can be written as 
\begin{equation}
\mathcal C_{\rm beam-space} \approx r\log_2 ({\texttt{SNR}}) + \sum_{i=1}^{r} \log_2\left(\frac{{\mu}^2_i}{r}\right) {\text{bits/s/Hz.}}
\end{equation} 
 As a result,  we can see that the receive SNR can be improved without extra power consumption while multiplexing gain could also be achieved by beam-space multiplexing. }


\vspace{-0.3cm}
\subsection{\textbf {Comparison of CSI Acquisition Between TDD and FDD }}
In order to align beam to each spatial data piece of beam-space multiplexing accurately, CSI should be acquired at the BS before the transmission. In FDD systems, CSI can be obtained through reference signal training and quantized feedback. However, the performance is limited by several non-ideal factors, such as feedback delay, quantization error, and imperfect CSI estimation at user side.  This will lead to the engineering implementation challenge due to heavy overhead of real-time feedback. The limitation becomes extremely severe in 5G, where the number of antennas in the BS is large and the
reference signal symbols are limited. The required independent
reference signal symbols are proportional to the number of antennas at the BS. The number of available reference signal symbols is fundamentally limited by the channel coherence time. Thus, the reference signal symbols will become insufficient for accurate channel estimation in FDD systems with large number of antennas at the BS.

Recently, compressed sensing (CS)-based channel estimation methods have been proposed to reduce the reference signal overhead for FDD massive MIMO systems, e.g., orthogonal matching pursuit (OMP) and least absolute shrinkage and selection operator (LASSO) algorithm \cite{Yangperformace}. \begin{figure}[!hbt]
	\centering
	\includegraphics[width=0.5\textwidth]{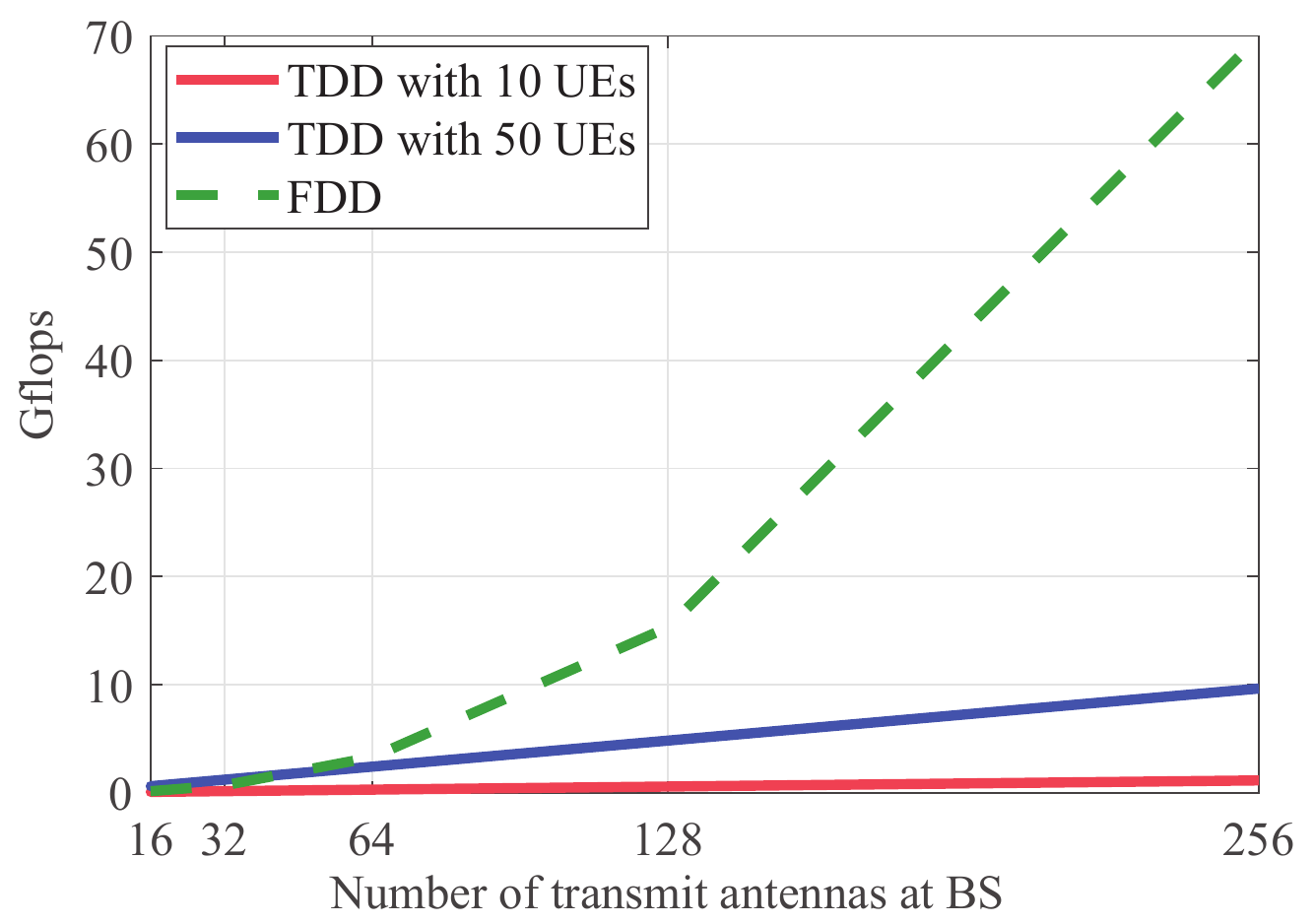}
	\caption{Computational complexity of CSI estimation for TDD and FDD systems.}
	\vspace{-0.4cm}
\end{figure}\label{Fig:2}
It has shown that the massive MIMO channel usually exhibits abundant sparsity in certain transformed domains due to limited local scatters in physical environments. However, there are some drawbacks for the CS-based algorithms. For example, group sparsity and other sparse structure in the different domains are not well exploited. This is because of the limitation of the solution for the CS optimization problem. 
{To overcome this problem, a structured compressing sensing (SCS)-based spatio-temporal joint channel estimation scheme is proposed to reduce the required pilot overhead, whereby the spatio-temporal common sparsity of delay-domain MIMO channels is leveraged. Moreover, the design of non-orthogonal pilot for CS-based adaptive CSI acquisition for improving channel estimation performance. }

\begin{figure*}[hbt]
	\begin{center}
		\includegraphics[width=1\textwidth]{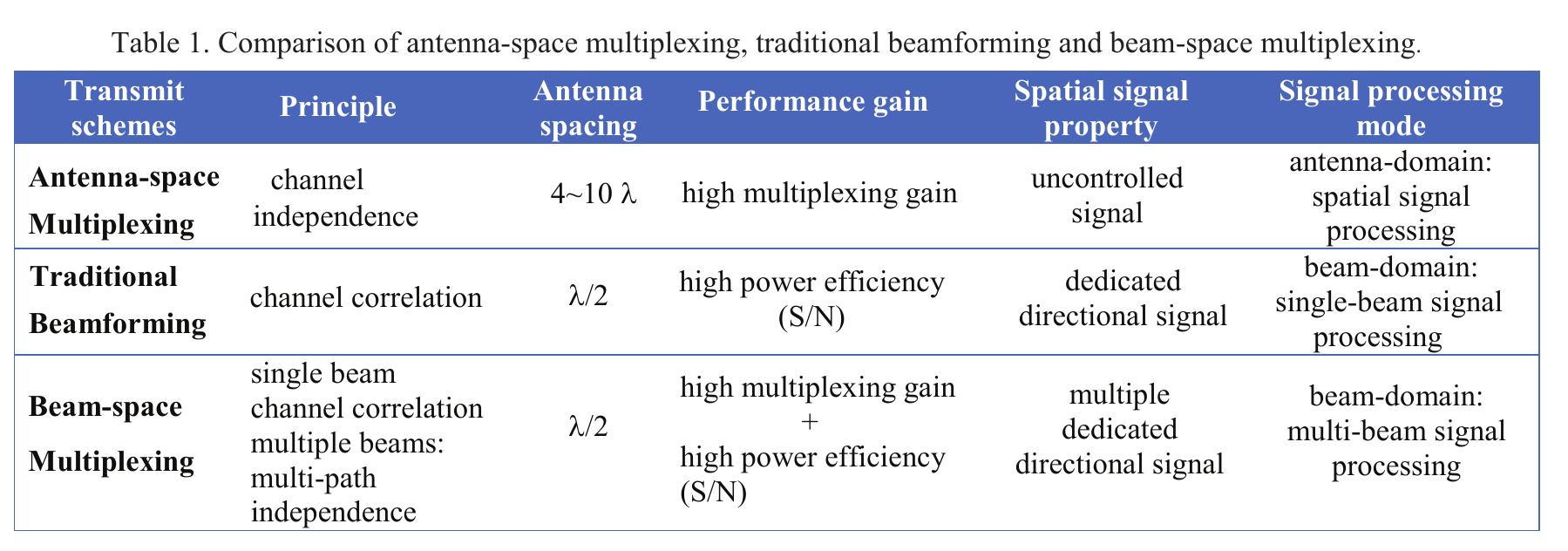}
	\end{center}\vspace{-0.2cm}
\end{figure*}

Alternatively, in TDD systems, channel reciprocity property can be utilized for CSI acquisition. The pilot overhead for CSI estimation in TDD massive MIMO systems is only proportional to the number of users.  Furthermore, as shown in Fig. 2, in contrast to FDD systems, the computational complexity for CSI estimation in TDD system is relatively much lower \cite{HanComp}. However, strictly synchronization and antenna calibration are required to guarantee the channel reciprocity. This is because reciprocity holds only in air channel between the BS and user side. If the transceiver of BS and users are considered, downlink and uplink channels are no longer reciprocal. Thus, in practice, calibration is needed to
restore channel reciprocity. Recently, user-assisted calibration is an enabling technology for TDD massive MIMO in 5G, which yields fast calibration processing with low overhead and high calibration accuracy. It is expected that more efficient calibration algorithms can be emerged such that the multi-fold gain of massive MIMO in TDD network can be fully realized.

\vspace{-0.3cm}
\subsection{ \textbf{Comparison of Engineering Implementation}}

\begin{figure}[!hbt]
	\centering
	\includegraphics[width=0.5\textwidth]{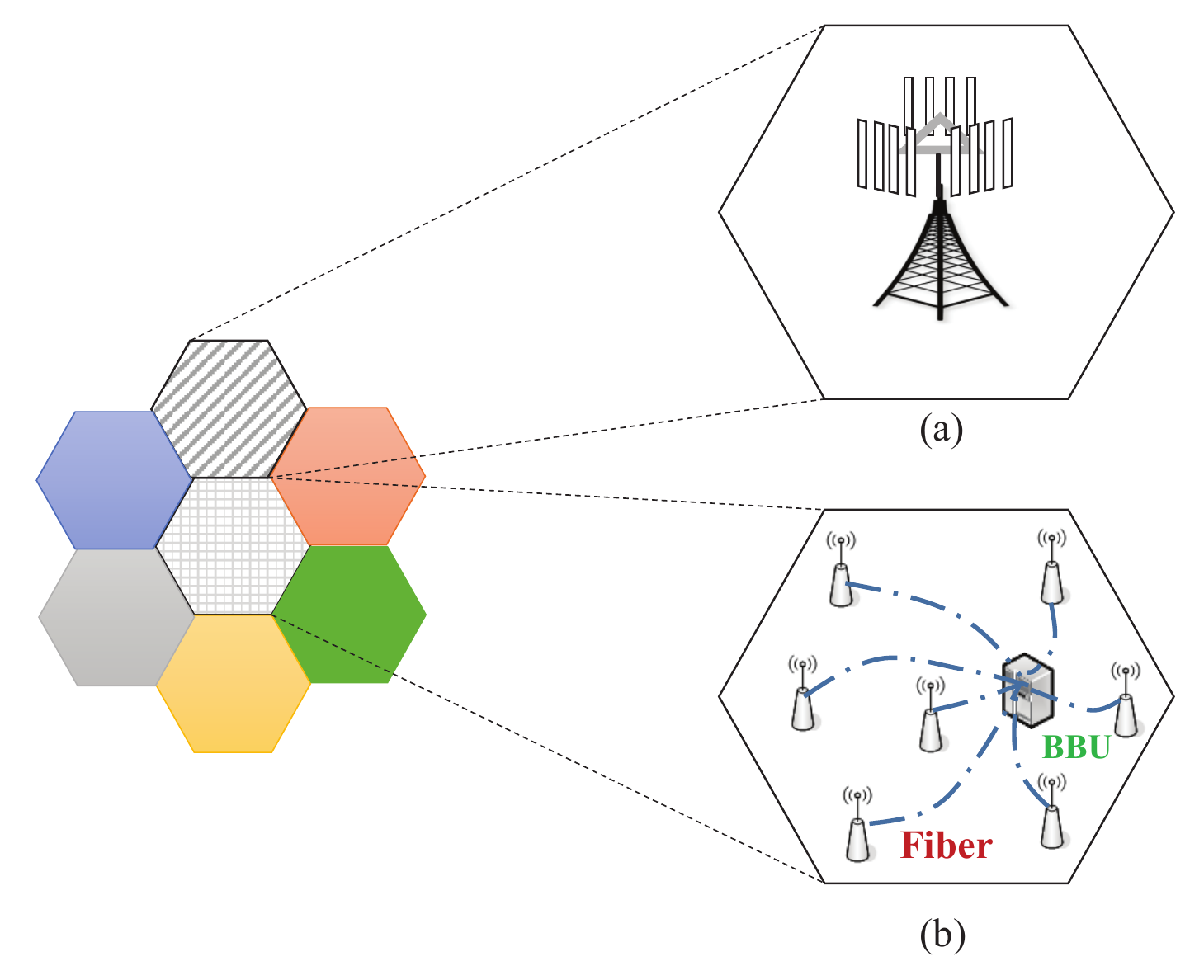}
	\caption{BS with different antenna deployment, (a) compact antenna array and (b) distributed antennas.}
	\vspace{-0.4cm}
\end{figure}\label{Fig:3}

We compare the beam-space multiplexing with the traditional antenna-space multiplexing by considering the practical engineering implementation constraints. For practical implementation with antenna-space spatial multiplexing, large-spaced (i.e., $4\!\sim \!10\lambda$) two and four antennas were used to provide enough decorrelation of multiple antenna signals, which facilitates multi-stream transmission for cell-center users. However, for cell-edge users, it will fall back to spatial diversity. From the engineering point of view, the number of BS antennas will be limited to a practical figure. If we want to further increase the number of antennas, two possible antenna array architectures of the antenna-space multiplexing are illustrated in Fig. 3:

\noindent {\underline{\it {(1) Compact antenna array:}}} Several antennas within one large panel or multiple antennas installed independently with different poles alternatively. A large panel will result in a large windward area and heavy weight, which is difficult to satisfy {the engineering constraints: weight should less than 47 Kg, windward area should less than 0.6 $\rm m^2$ (Source: China Tower).} Multi–poles solution makes it difficult to be installed on an ordinary tower.  Hence, in practical, the number of antennas is limited to $4 \!\sim \! 8$, and the inter-antenna distance usually reduced from $4 \lambda$ to $1\!\sim \! 2 \lambda$. This will obviously decrease the spatial multiplexing performance. 

\noindent {\underline{\it {(2) Distributed antennas:}}} A theoretical architecture of massive distributed antennas is illustrated as many small sites scattered within the cell area, each site with one or few antennas, so the
distance between different antennas is far than $10 \lambda$. Centralized baseband processing is accomplished through ideal fronthaul fibers. This architecture can give a promising performance, but extremely difficult in engineering. Both hard for site acquisition, distributed antenna calibration, fiber laying, and distributed massive MIMO architecture will suffer from high installation and maintenance costs.

On the contrary, beam-space multiplexing makes it easier. The antenna space is usually $\lambda/2$, large number of antennas can be realized within a practical panel size. In addition, as higher carrier frequency is introduced into 5G system, cell radius gets much smaller, the probability of the user located at the interference limited/cell edge area is greatly increased. Beamforming can improve the SNR for cell-edge users by high array gain, which may still keep beam-space multiplexing.

{In summary, beam-space multiplexing is a feasible technical scheme and innovation under engineering constraints (volume, power consumption, weight, cost and implementations, etc.). It has solved three main scientific and engineering problems of using MIMO technologies in current wireless communication system: improvement of SE, expanded coverage, and feasible engineering deployment.   Therein, power consumption, cost and computational complexity are closely related. }

\begin{figure*}[!htbp]
	\centering
	\subfigure[]{
		\begin{minipage}[h]{0.48\textwidth}
			\centering
			\includegraphics[width=1\textwidth]{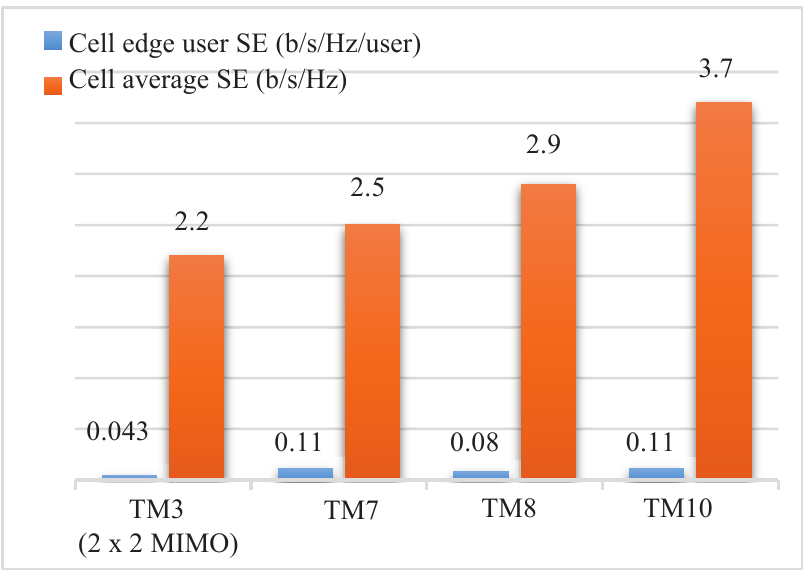}
		\end{minipage}%
	}%
	\subfigure[]{
		\begin{minipage}[h]{0.5\textwidth}
			\centering
			\includegraphics[width=1\textwidth]{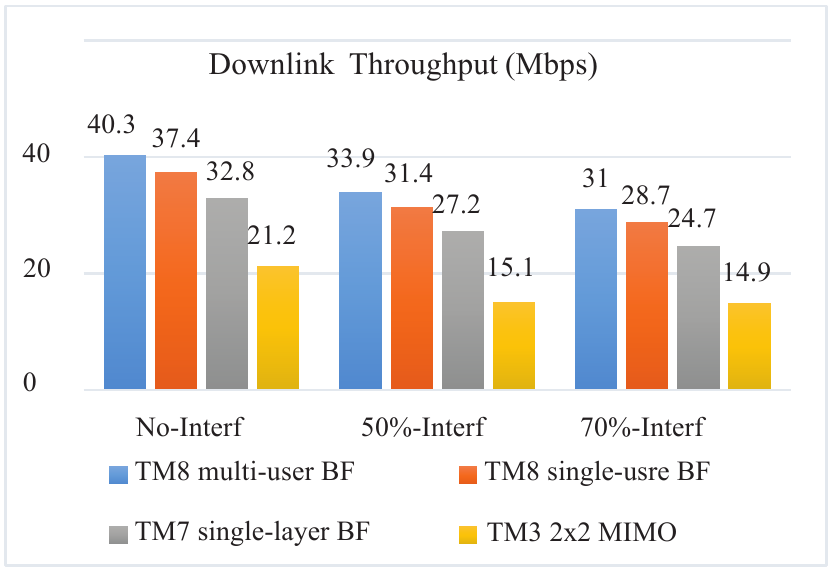}
		\end{minipage}%
	}%
	
	\centering
	\caption{Downlink performance of different transmission schemes in TD-LTE:  a) SE of cell average and cell edge user, and (b) downlink user throughput.}
		\vspace{-0.4cm}
\end{figure*}

\vspace{0.8cm}
\section{Beam-space Multiplexing in TD-LTE: Multi-layer Beamforming}

3GPP TD-LTE is the first mobile communication system adopts the beam-space multiplexing, in which multi-layer beamforming is exploited for simultaneously multiplexing multiple data layers and improving the link quality.
\subsection{\textbf {Key Technologies}}
With controllable spatial distribution of transmitted signal, beamforming can be used to enhance transmission by focusing the radiation in desired direction. As the equivalent channel quality is improved after beamforming, the data rate can be boosted as well with higher modulation-and-coding schemes (MCSs). The improvement in SINR and the extensive application of dual-polarization antenna arrays are also in favor of the transmission of more spatial layers.

In multi-layer beamforming for TDD system, with full CSI at transmitter side, an optimized beam can be formed by adjusting the weights applied on the array to match the eigenvector of each layer.  Therefore, the peak data rate for users with favorable channel characteristics can be significantly improved. 

From system-level perspective, if the system load is high enough, it is naturally to allocate a group of users with the same 
set of time-frequency resources and differentiate them with different beams in spatial domain. In multi-user multi-layer beamforming, an overall optimization in scheduling, beamforming and adaptive modulation and coding (AMC) can be achieved at system level. Consequently, the gains of multi-user time-frequency selective scheduling, beamforming as well as spatial multiplexing can be obtained by the system.

Besides the capability of enhancing spectrum efficiency for each cell independently in single-cell operation, beamforming technology also offers a flexible way of mitigating interference among cells. That is, through network-level coordination, a cluster of cells can optimize the beamforming and scheduling jointly to minimize the inter-site interference. What’s more, the antenna array of multiple site can be used jointly to form beams to further enhance the transmission.    

With the introduction of active antenna system (AAS) technology, it is possible to have more controllable antenna ports as baseband. Therefore, higher spatial degree-of-freedom are available for implementing more flexible beamforming technology. For example, for a planner array, the beam can be formed in both horizontal and vertical domains to fully match the actual spatial propagation properties of signal in 3D space. In practical, such 3D beamforming technology can be used to 
enhance the outdoor-to-indoor coverage of high-rise buildings. For cell edge users, higher flexibility can be used for inter-cell interference mitigation or coordination. For single-user case, each user can be served with more accurate beams, and hence higher channel quality can be achieved. For multi-user transmission, with higher spatial resolution, more users are possible to be co-scheduled to improve the system SE. The inter-user interference can also be better mitigated with such 3D beams. 

\vspace{-0.2cm}
\subsection{\textbf {Standardization}}
Beam-space multiplexing was firstly introduced in LTE Rel-9 to support single-user dual-layer transmission (transmit mode 8, TM8) with two beams, as well as multi-user multiplexing. Furthermore, LTE Release 10 can support up to eight layers of beam-space multiplexing, with further enhances 
of the multiplexing gain, and being called TM9. It is designed to help the interference minimization between base stations, and to maximize signal stability, as well as to boost performance. The new TM9 enables the enhancement of high network capabilities and performances with the minimum added overhead. TM9 is designed to combine the advantages of 
SE (using higher order MIMO) and the higher data rates of cell-edge users, the coverage and interference management (using beam-space multiplexing). Note that the flexible switching among single-layer, multiple-layer, single-user, multiple-user, and beam-space multiplexing schemes is possible within TM9. Beam-space multiplexing for CoMP transmission (TM10) in multi-cell was 
also standardized in LTE Release 11, targeting the mitigation of not only intra-cell interference, but also inter-cell interference \cite{ChenADPT}. The 3GPP LTE Release 12 further optimizes multi-antenna technology including enhancements of the downlink four-antenna transmission technology and CoMP technology, and studies in two-dimensional (referring to horizontal dimension and vertical dimension).

\vspace{-0.3cm}
\subsection{\textbf {System Evaluation and Field Results }}
It has shown that the benefits of beam-space multiplexing in the link level. In this section, we present numbers of results from system-level simulations and performance of running commercial networks. Firstly, beam-space multiplexing schemes are evaluated by system-level simulation platform. The simulation is carried out by software written in C/ C++. The ITU urban macro (UMa) scenario is constructed with 19 BSs, each BS covers 3 sectors, and 10 users are uniformly dropped in each sector. Each sector is equipped with 8 cross-polarized transmitting antennas, and there are 2 receiving antennas at a user. The bandwidth of 10 MHz at 2.6 GHz frequency is assumed. 

During the evaluation, single-user transmission is assumed for TM3, TM7 and TM8 in a single cell network while maximum 4 users/cell could be scheduled for TM10 in a homogeneous network with three co-located cells sharing the same BS. In addition, multiple-user transmission is enabled. 

For LTE TM3, $2 \times 2$ MIMO open loop spatial multiplexing was adopted. It is evident from Fig. 4 (a) that TM7 (single-layer beamforming) can achieve 156\% SE gain for cell-edge users. This is because TM7 provides link reliability due to beamforming gain. TM8 combines spatial multiplexing and beamforming, enabling multiple layers of transmission in the beam-space domain while retaining beamforming gains.  

As shown in the Fig. 4 (a), there is 16\% performance gain of cell average SE for TM8 over TM7 with single-user transmission. Coordinated beamforming with LTE TM10 is also evaluated. It is shown that TM10 provides significant gain for cell average SE users as well as SE for cell-edge users. With coordinated beamforming in multiple cells, judiciously selected user and beamforming weight can lower interference. 

Moreover, we present test results of different multi-antenna transmission schemes in the commercial TD-LTE networks of China Mobile. The test case was carried out in a densely populated urban area with 20 MHz bandwidth {with frequency 2575 \!$\sim$\! 2595 MHz, where users are randomly distributed in a specific cell. To model the different interference loads in a specific cell, interference signals were transmitted from neighboring cells with 0\%, 50\% or 70\% time-frequency resource reuse from the interesting cell. The test is traversing a continuous coverage area consisting of 19 BSs. } As shown in Fig. 4 (b), TM7 performed much better than that of TM3 over all interference load. In particular, users may experience low SINR with 50\% and 70\% interference load and TM3 may not be able to support spatial multiplexing transmission. Then, multi-layer beamforming was compared to single-layer beamforming by evaluating LTE TM8 and TM7 with single-user transmission. From the test results, 15\% cell throughput gain can be achieved by adopting multi-layer transmission. Lastly, to demonstrate the benefits of multi-user transmission, LTE TM8 with/without multi-user transmission is evaluated. From the test results, there is $5\sim 10 \%$ performance gain being achieved with multi-user transmission.

\vspace{0.8cm}
\section{Beam-space Multiplexing in 5G: Massive Beamforming}
The 5G NR air interface is expected to satisfy the requirements of three deployment scenarios: enhanced mobile broadband (eMBB), massive machine type communications (mMTC) and ultra-reliable and low latency communications (URLLC). The performance gains in terms of system capacity, user experience as well as spectral and energy efficiency with massive MIMO were proven by theoretical researches in \cite{Tom}. 
Based on that theory, with the increase of antenna number, the channels of different users tend to be orthogonal to each other. Consequently, furthermore, as the beam becomes narrower and the gain is higher, the inter-user interference and additive noise will vanish if the number of antennas approaches to infinitely. Therefore, beam-space multiplexing with massive MIMO, which is termed massive beamforming, is expected to be the mainstream multi-antenna technology supporting the continuous improvement of TDD competitiveness, which tends to be the global 5G mainstream standard \cite{R1-164256, RR1-166478}.

\subsection{\textbf {Key Technologies}}
The results from early researches and early evaluations exposed the potential of massive beamforming in boosting spectrum efficiency for the 5G system. There are several challenges in its standardization and practical application:

\noindent{\underline{\it (1)   Channel modeling and deployment scenarios:}} The performance of multi-antenna system highly relies on the actual deployment scenario. Therefore, the accurate modeling of massive beamforming channel and the deployment scenarios are of great importance to evaluate the performance of massive beamforming algorithms and make comparison of standardization schemes. When the number of antenna elements is high, the channel model becomes more complicated to reflect the realistic properties of antenna arrangements and antenna correlation. Facing to the restriction on array size, massive beamforming is expected to be more suitable for higher frequency bands, i.e., above 6 GHz. The typical deployment scenarios and the corresponding channel model should be identified based on field measurements, especially for frequency bands above 6 GHz. In 3GPP, based on large-scale field measurements, the channel and scenario models for massive beamforming with up to 100 GHz frequency band have already been specified. 

\noindent{  \underline{\it (2) Transmission/reception schemes:}}  With increased scale of antenna array, larger number of users and wider bandwidth, more complicated high-dimension matrix computations should be conducted in transmission, reception, detection, and scheduling for massive beamforming system.  In system design, the trade-off between performance and overhead/complexity should always be considered including aspects such as reference signal, feedback mechanism, control signaling, broadcast, initial access, etc. Generally speaking, the following four categories of issues are the most essential aspects for massive beamforming system design: beamforming scheme, channel state information measurement and feedback, coverage enhancement and high-mobility solution, multi-user scheduling and resource management. 

\noindent{  \underline{\it (3) Design of Large-scale active antenna array:}} Besides the emerging theory supporting spatial extension of MIMO transmission, the evolution of antenna system and the enhancement of processing capability of more advanced devices shall also provide solid grounds for implementing massive beamforming technology in practical deployment scenarios. For instance, the passive antenna systems, which have been extensively used in current wireless access networks, are not suitable to support further extension of spatial dimension, especially in vertical domain.  On the other hand, with the integration of radio frequency (RF) and partial or even full baseband functionalities into the antenna system, AAS provides a more feasible platform for the network to control more antennas in both horizontal and vertical domains. Therefore, as promised by massive MIMO theory, the more flexible beamforming transmission/reception can be achieved for more precisely matching of 3D channel properties of larger number of users in practical system.

\subsection{\textbf {Standardization}}
Related standardization works in 3GPP started with study items on full-dimension MIMO (FD-MIMO), an initial version of massive MIMO based on planar antenna array of AAS. In Rel-13, a study conducted for MIMO enhancements to extend the current support from 8 up to 64 transmit antennas was conducted, thus targeting a massive number of controllable antenna elements at the BS. The study considered simultaneous horizontal and vertical adaptive transmission, utilizing 2D antenna arrays with both closed loop and open loop (beamforming) operation modes. Enhanced feedback and cell sounding for both closed and open loop operations offer significant performance improvement for both single-user and multi-user MIMO (SU-MIMO and MU-MIMO), which led to standard enhancements for up to 16 antennas in Rel-13.  In Rel-14, the MIMO evolution will continue, targeting up to 32 transmit antennas. Currently, both SU-MIMO and MU-MIMO, the performance limitation is related to the quality of the channel knowledge at the transmitter. It motivates the investigation of new feedback methods for high-resolution feedback and for the existing precoding codebook-based scheme. The challenge is how to get sufficiently good transmitter channel knowledge for FDD and TDD when full reciprocity is not available, for instance when the terminal has less transmit antennas than receive antennas.

In 5G NR Rel-15, massive-MIMO antenna systems are extended to support a maximum of 256 antenna elements for sub-6 GHz frequencies. For future deployment of NR system in mmWave band, an active antenna array with tens to hundreds of elements can be allocated in inches, which efficiently take the advantage of massive beamforming. Targeting higher requirement on system performance and more flexible deployment scenarios, the design and standardization of massive beamforming technology encounter more challenges in 5G. Hence, the following aspects need to be considered:

\noindent { \underline{\it (1) Frequency band}}

Due to the lack of frequency resources below 6 GHz, further extension of frequency band is one of the most important ways for NR system to achieve the requirement for 5G. In Rel-15, up to 52.6 GHz can be supported, whereas in the upcoming releases, up to 100 GHz frequency bands will be considered. In higher frequency bands, more spectrum resources are available. However, the propagation characteristics such as path loss, blockage, penetration loss, and so on, tend to be more severe to wireless communication system. In such case, massive beamforming serves as an important manner to overcome the disadvantages of propagation environments and offer flexible coverage to the system in mmWave bands. 

With the increase of carrier frequency, the size of antenna array can be reduced accordingly. Or, for the antenna array with the same size, more antennas can be used. In the other words, the extension of MIMO dimension is also benefited by the increase of carrier frequency. On the other hand, considering the cost, power consumption and complexity, it’s hard to implement full-digital antenna array in mmWave bands. Therefore, a more reasonable solution is the so-called hybrid antenna array. Based on such a structure, a two-step beamforming, or hybrid analog and digital beamforming has to be used. In analog domain, as the channel cannot be estimated for each pair of antennas via reference signal inserted in digital domain, the mechanism for beam searching, tracking, reporting and recovery need to be established. In the standardization of NR system, the aforementioned schemes were categorized as beam management and beam failure recovery.

In order to compensate possibly large path loss, high-gain beams are usually used in mmWave bands. Compared with sub 6 GHz, the use of narrower beams in higher frequency bands also limits the effective spread of rays in both angular and time domain. Consequently, the selectivity properties of beamformed equivalent channel are changed as well. Therefore, the selection of system parameters regarding factors such as beamforming and scheduling granularity are reconsidered for above 6 GHz bands. Besides, some other issues also need to be considered. For example, the impact of phase noise on demodulation and more realistic studies of mmWave mobility support and system evaluation and field trial need further investigations.

\noindent {  \underline{\it (2) Enhancement of multi-user MIMO}}

Although there are abundant frequency resources available in mmWave bands, considering the increasing of users and traffic load in the future, improvement of spectrum efficiency is still of great importance to 5G system. With large-scale antenna array and more flexible 3D beamforming capability, massive beamforming can be used to further enhance the performance of multi-user transmission. Compared with single-user beamforming, the performance of MU-MIMO relies heavily on the accuracy of beamforming and scheduling based on CSI feedback. To that end, channel reciprocity can be used for the network side to acquire CSI precisely for TDD system. For the case there is no channel reciprocity doesn’t hold, codebook-based feedback mechanisms can still be used. In NR, two types of codebooks were adopted. Namely, the Type I codebook for ordinary CSI feedback, e.g., for SU-MIMO, and higher-resolution Type II codebook is optimized for MU-MIMO. Depending on the configuration of Type II codebook, the overhead for precoding matrix indicator (PMI) reporting could be rather large. The trade-off between overhead and performance gain should thus be considered carefully in actual implementation of such mechanism.

\noindent {  \underline{\it (3) Flexibility}}

Aiming at more flexible deployment scenarios and diverse kinds of traffics, the design and standardization of massive beamforming schemes for 5G should full consider the flexibility in parameter configuration and system operation. As shown in the following examples, such requirements on flexibility were reflected in aspects such as reference signal and CSI feedback framework designs in NR.

For the sake of forward compatibility and power consumption reduction, the always-on signals like cell specific reference signal (CRS) in LTE are avoided as much as possible. Depending on the actual usage, CSI-RS can be configured to transmit in given time-frequency resources. The number of ports and the purpose of CSI-RS are all configurable. In addition, the desired CSI-RS pattern can be obtained based on the aggregation of basic patterns.

To reduce the decoding delay and to meet the requirement of certain low-latency traffic, the first symbol of demodulation reference signal (DM-RS) in time domain is located to be close to the beginning of scheduling duration as much as possible. For different mobility, 1, 2 or even 3 sets of additional DM-RS symbols can be configured. Similar design is adopted for both downlink and uplink to better overcome inter-link interference between different links. Two types of DM-RS with different 
densities and the number of orthogonal ports can be configured for different scenarios. A unified feedback framework is introduced in 5G NR, based on which both CSI and beam measurement related information can be reported. With such a 
framework, according to the desired manner of feedback, parameters can all be configured including the reference signal for channel and interference measurement, type of CSI feedback, codebook type, time domain property (periodic, aperiodic, semi-persistence), and feedback granularity, etc.

\subsection{\textbf {System Evaluation and Field Trial }}

\begin{figure*}[htbp]
	\centering
	
	\subfigure[]{
		
	\begin{minipage}[b]{0.5\textwidth}
		
		\includegraphics[width=1\textwidth]{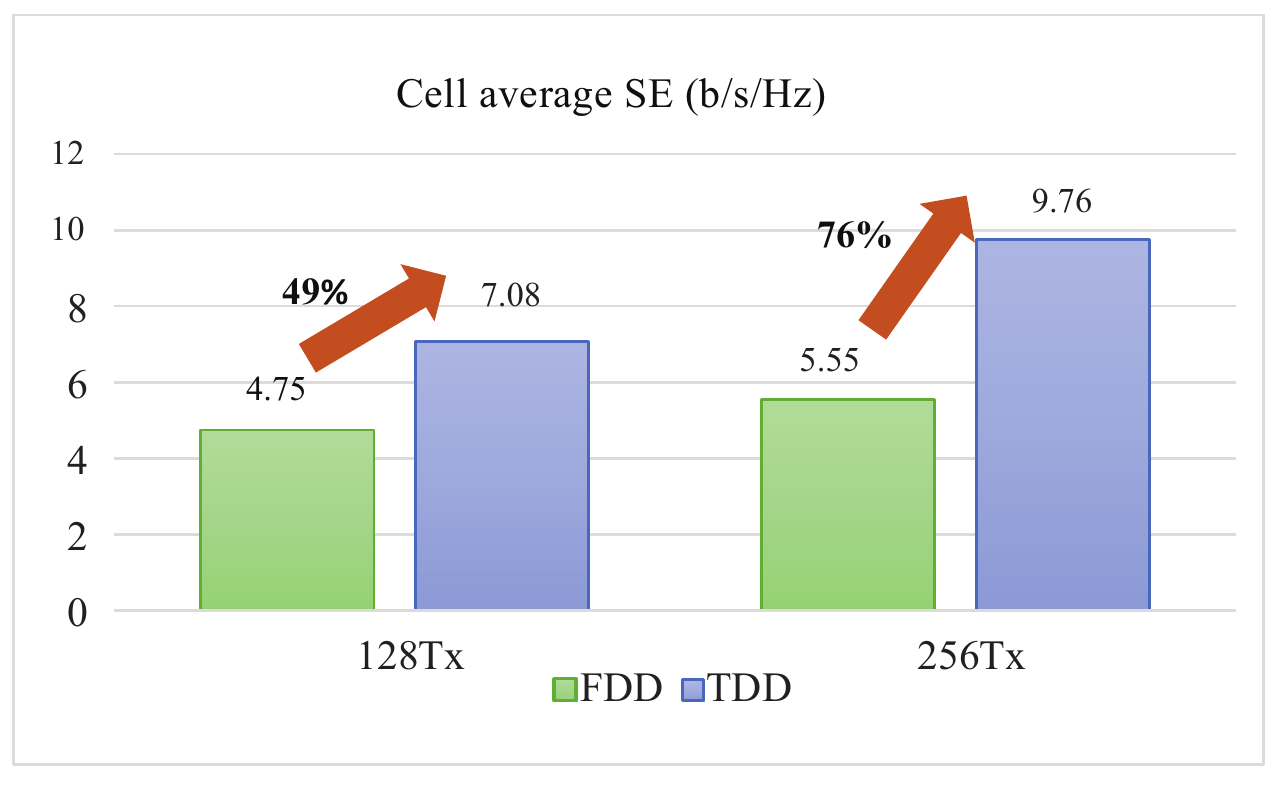}
		
		\end{minipage}%
	}%
	\subfigure[]{
			
		\begin{minipage}[b]{0.5\textwidth}
			
			\includegraphics[width=1\textwidth]{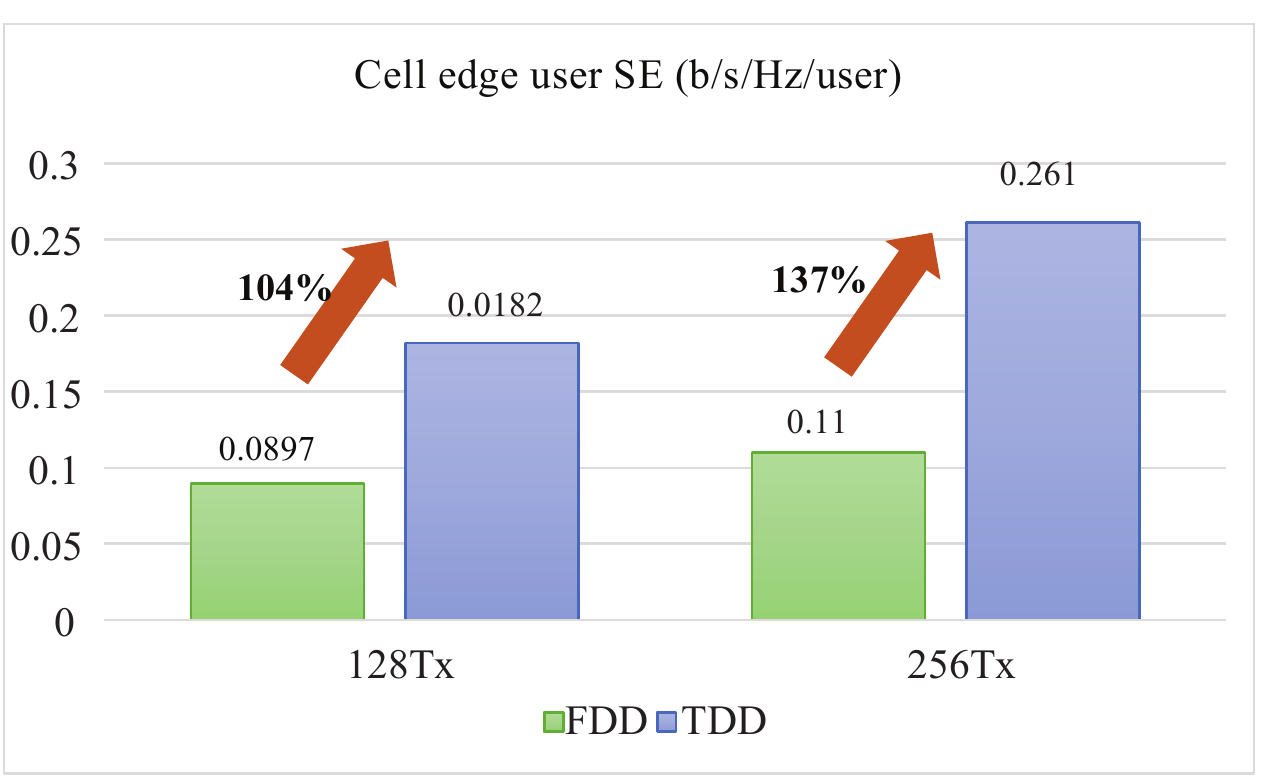}
			
		\end{minipage}%
	}%

	\subfigure[]{
		\begin{minipage}[b]{0.5\textwidth}
		
		\includegraphics[width=1\textwidth, height=3in]{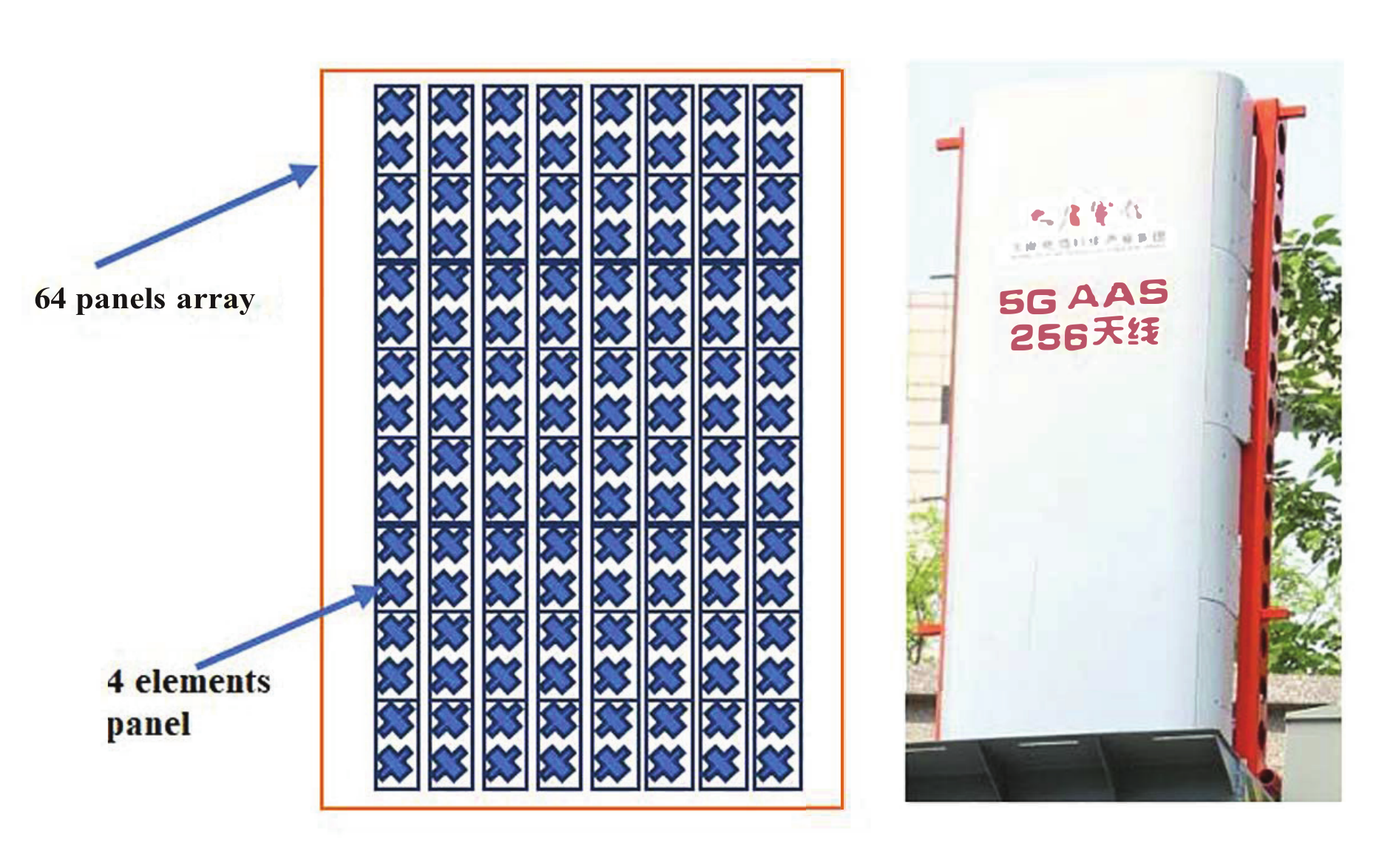}
		
		\end{minipage}
	}%
	\subfigure[]{
			\begin{minipage}[b]{0.5\textwidth}
			
			\includegraphics[width= 1\textwidth]{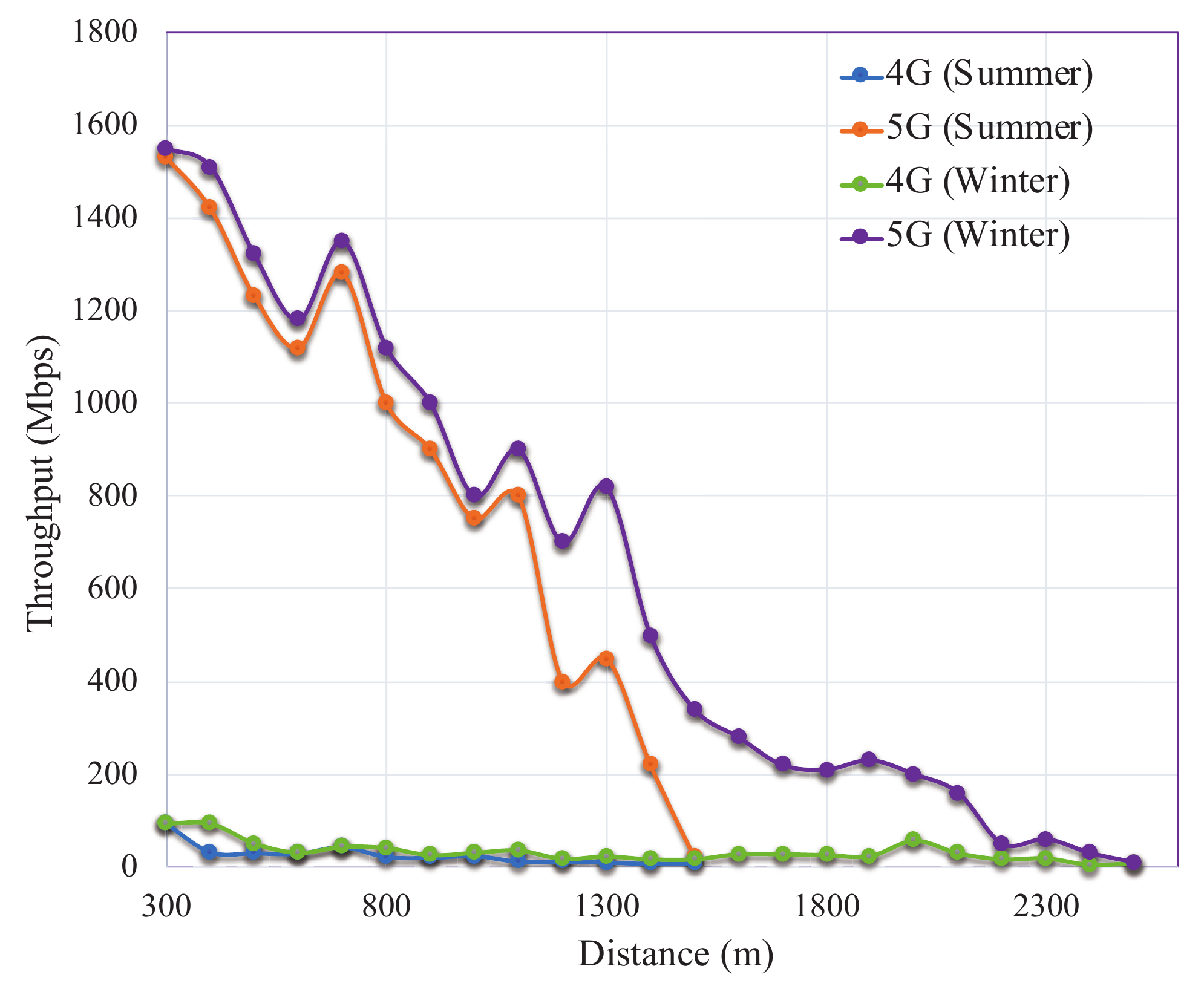}
			
		\end{minipage}
	}%
	
	\centering
	\caption{System evaluation and field trial of massive beamforming: a) and b) the performance of massive beamforming schemes in FDD and TDD; c) an example of a 256-element active antenna array (prototype testing with 256-antenna elements array and with channel reciprocity at 3.5 GHz frequency); d) comparison of throughput performance of 5G massive beamforming and 4G LTE system.} \label{fig:5}
	\vspace{-0.4cm}
\end{figure*}

Based on the channel model developed in 3GPP, an evaluation of massive beamforming was carried out in both TDD and FDD. As shown in Fig. 5 (a) and Fig. (b), massive beamforming based on channel reciprocity in TDD provides significant gain over massive beamforming based on feedback in FDD. With 128 antennas, 49\% and 104\% gain are achieved in terms of spectral efficiency of cell average and cell edge, respectively. The gain rises to 76\% and 137\% when 256 antennas are used. The more the antennas are equipped, the larger the gain is observed. It should be noted that massive beamforming used in both TDD and FDD in 5G system. Utilizing channel reciprocity in TDD alleviates the problem of CSI acquisition. This makes TDD a more promising candidate for 5G system with large antenna array.

Besides, Fig. 5 (c) shows a prototype of active antenna array with 256 elements developed by CATT/Datang in 2016, where 4 elements are grouped to form a panel and 64 panels form a panel array. Each panel is mapped to two TXRUs, one on each polarization.  At the first phase, it supports 20 parallel streams transmission such that peak data rate could exceed 4 Gbps, and the upgraded one could support 32 streams with the peak transmission rate of 6 Gbps. Compared with 4G technology, it can achieve 5~10 times spectrum efficiency improvement.

In Fig. 5 (d), the field test results from 5G massive beamforming and 4G LTE system are shown. The carrier frequency of 5G massive beamforming and 4G LTE system are 3.5 GHz and 2.6 GHz, respectively. The system bandwidth is 100 MHz, while 20 MHz bandwidth is used for 4G LTE system. The active antenna array used for 5G massive beamforming is equipped with 64 digital channels and 192 antenna elements. For LTE system in the field test, an 8-antenna array is used for each site.

Based on the test results, it is observed that 5G massive beamforming outperforms 4G LTE system significantly in most part of the interested coverage area. The peak throughput as high as 1.6 Gbps can be obtained for users in the most advantageous locations. With higher carrier frequency, the system performance of 5G massive beamforming is more sensitive to the distance. Therefore, the change of throughput curves for 5G massive beamforming tend to be steeper compared with that of 4G LTE system. However, even for coverage distance of 1000m, around 50\% performance gain can still be achieved for 5G massive beamforming over 4G LTE system.  In addition, as the foliage loss increases with higher carrier frequency, the performance of 5G beamforming shows higher dependency to the season. As consequence, a larger difference of throughput can be observed with the increase of coverage distance in summer and winter for 5G beamforming.

\vspace{0.8cm}
\section{Future Research Directions of Beam-space Multiplexing in 6G and Beyond}
With the great efforts of global researchers in academia and industry, 5G is at our fingertips. Since 2019, several countries and cities have started to deploy commercial 5G cellular networks. Therein, massive MIMO, one of the key technologies  in 5G is now mature. The key ingredients of massive MIMO have been standardized in Rel-15 \cite{Emil}. It is, therefore, time for MIMO researchers to change focus towards future research directions of beam-space multiplexing in 6G and beyond.  6G will go beyond mobile Internet and will be required to support ubiquitous Artificial Intelligence (AI) services from the core to the end devices of the network \cite{KBL}. To this end, we outline four emerging beam-space multiplexing related research directions to fulfill the requirements for expected services in 6G and beyond: massive beamforming for XL-MIMO and LEO satellites communication, data-driven intelligent massive beamforming, and multi-target spatial signal processing, i.e., joint communication and radar sensing, etc.

\noindent {  \underline {\it (1) Massive Beamforming for XL-MIMO:}} As a straightforward evolution result of massive beamforming, the extremely large number of antennas with enhanced spatial signal processing (i.e., massive beamforming) will be a forward-looking research direction. We can expect a near future where hundreds or thousands of antennas are used to serve very many users, e.g., MTC devices. In general, there are three types of ways to create XL-MIMO: { 
\begin{itemize}
\item  {\textbf {Extremely Large Aperture Arrays (ELAA): }} Extremely large arrays of classic antennas distributed over a substantially larger area, typically embedded in the window or ceiling of a building of large dimension.

\item  {\textbf { Large intelligent surface (LIS):}}A limited surface area consists of uncountable infinite number of antennas. A possible implementation is with tightly coupled array of discrete active antennas or nearly passive reflecting elements but possibly other ways, i.e., meta-surfaces.

\item  {\textbf { Cell Free XL-MIMO:}} Hundreds or thousands of distributed BS antennas at distant geographical locations that are jointly and coherently serving many distributed users.
\end{itemize}
}

In XL-MIMO system, super spatial resolution and favorable propagation are expected to enable incredibly low transmit powers and unprecedented spatial multiplexing capabilities. It can provide orders-of-magnitude higher area throughput compared to the current massive MIMO with centralized compact antennas, which can practically deliver.

However, XL-MIMO is still in its original state, there are several challenging problems when employing beam-space multiplexing. 	Firstly, large array aperture extended near-field radium such that signals coming from widely different directions cannot be separated. Therefore, the principal problem in ELAA is to setup interference-rejecting precoding/combining schemes, such as zero-forcing (ZF) and minimum mean square error (MMSE) beamforming, in a distributed or hierarchical way. { Furthermore, beam squint effect occurs with extremely large number antenna at the BS, which means that different frequency would have different beam direction, i.e., frequency-selective beamforming \cite{Baoleiwang}}.  Besides, due to non-stationary spatial channel properties in ELAA, the design of dynamic beamforming schemes is indispensable. { The impact of spatial non-stationary characterized by visibility regions where the channel energy is significant on the portion of the XL-array.  The performance of beam-space multiplexing schemes for the downlink multi-user  needs further investigations. Further researches on channel modeling, system-level performance evaluation as well as practical implementation for ELAA are also highly required. }

\noindent {  \underline {\it (2) Massive Beamforming for LEO Satellite Communication:}} LEO satellite communication is another emerging research area in wireless mobile communication. With hundreds or thousands of satellites located at dozens of orbits usually 500~2000 kilometers above the earth, LEO satellite systems can provide broadband mobile services for both wide range coverage as well as vertical applications. Compared with previous geostationary earth orbit (GEO) system, LEO system shows advantages of much lower transmission delay and much lower path loss. Therefore, an integrated satellite and terrestrial global seamless coverage can be easily formed by integrating LEO satellite systems and terrestrial cellular system beyond 5G systems, promoting continuous extension for wireless mobile communication application.

{Similar to} the terrestrial system, massive beamforming technology is expected to be applied in the LEO system. For the service link between the terminal and the satellite, usually frequency band Ku, Ka or Q is used. Both the terminal and the satellite can use phased arrays antenna to send and receive massive beamforming for overcoming the propagation loss. { However, beams need to be dynamically scheduled for higher throughput or wider coverage.} Furthermore, massive beamforming can also form the type of directional beam, which could realize flexible communication with a particular user. {The narrower beam width leads to a higher antenna gain, as a result, the SE is increased. Furthermore, massive beamforming allows beams that are far apart to reuse frequency, which may bring about strong inter-beam interference due to the non-zero side lobes. Therefore, Therefore, side lobe suppression technologies are required for the use of massive beamforming.} In addition to massive beamforming between the terminal and the satellite, high frequency band such as mmWave and TeraHertz (THz) can also usually be used in the inter-link between satellites to realize high-speed data transmission, flexible network architecture and routing transfer applications. {However, accurate and fast estimation of angle-of-arrival (AoA) is the key enabling technology underlying wideband multi-beam LEO satellite communication systems.  It is vital for improving the beamforming gain and radio link quality, as well as increasing the data rate, especially in low SNR regions.
	Besides, the use of massive beamforming in hybrid satellite-terrestrial communication for Internet-of-Things (IoT) is an emerging topic, as the requirements and constraints for IoT connectivity are significantly different from the broadband connections.
}

\noindent {  \underline {\it (3) Data-driven Intelligent Massive Beamforming:}} Machine learning (ML) has been applied to a large variety of problems in wireless communication networks, including network management, self-organization, self-healing and physical layer optimizations (see [13] and references therein).  More and more trails show that ML can improve the performance of existing model-based algorithms with reduced implementation complexity. It is also opening a door for some use cases, which the characteristics of system are hard to model or analyze by conventional way yet can be learned from data. For example, the use of deep learning (DL) can significantly improve the complexity-performance trade-off of power allocation, compared to traditional optimization-oriented or model-based methods. For instance, DL can be used for channel estimation and beam-selection in massive MIMO systems. {Besides, the existing work has shown that  BS can find optimal beamforming vectors from a beam steering-based codebook  in  a mmWave communication system through deep reinforcement learning (DRL). Nevertheless, if the size of beamforming codebook is very large, it is impossible to find the optimal beamforming vectors by traditional model-based method.}

However, ML-empowered intelligent massive beamforming is still in its infancy. Several major obstacles need to be conquered. The first urgent issue is how to acquire data set for training the ML model and identifying the right use cases for ML in intelligent massive beamforming design. The lack of freely available data sets impairs the data-driven lines of investigation. Then, it is challenging to implement learning algorithms for massive beamforming with limited computation
capacity of radio hardware in real time. ML inference with hard real-time constraints at the micro-to nanosecond timescales is required. 

\noindent {  \underline {\it (4) Multi-target spatial signal processing: }} In addition to mobile broadband communication services, the massive antennas with high spatial resolution { can be used for the non-communication applications, such as radar, environment sensing and positioning.  With large-scale arrays, the system performance of target detection, parameter estimation, and interference rejection can be improved by high the spatial diversity gain and spatial resolution.  As a result, the number of targets in radar applications that can be uniquely identified from spatially multiple correlated or uncorrelated probing signals. Furthermore,} the property of  high directivity steering beams in massive MIMO systems enables, in principle, superior positioning precision. In particular, the combination of massive beamforming, high temporal resolution can be achieved, and then accurate range measurements can be obtained. One major potential application for joint communication and radar sensing is in vehicular networks, where signals can also be used for sensing the environment for object detection and collision avoidance. It is also very meaningful in mobile networks with other simultaneous communication service, such as drone’s detection and position at the airport.

Nevertheless, there are many challenging problems and possible improvements yet to be done.  { The main challenges is the robust beamforming and waveform design for target detection and estimation, as the echoes produced from nearly all surfaces (e.g., ground, sea, and buildings) when illuminated by a massive MIMO radar.} The challenge has a scalable complexity with the number of antennas and is implementable in a centralized or distributed manner based on the antenna array configuration. Furthermore, we require massive beamforming vector generation with quantized magnitude and phase values. We also require communication and sensing sub-beam combination methods optimized with respect to certain criterion. Besides, sensing algorithms that work for high-dimension and off-grid models that can resolve AoA and angle-of-departure (AoD) beyond the conventional way of scanning need further investigation as well. {Moreover,} some other problems need to be solved such that massive MIMO positioning can be realized, including channel models for positioning at different frequencies, signal processing and positioning algorithm, as well as engineering issues, e.g., synchronization and antenna positions calibration.

\vspace{0.8cm}
\section{Conclusions}
{This article has presented a comprehensive overview of beam-space multiplexing from engineering and theoretical perspectives. Firstly, we clarified the fundamental theory of beam-space multiplexing including theoretical analysis, CSI acquisition, and engineering implementation aspects. Then, we summarized the 3GPP standardization of beam-space multiplexing in 4G as well as the ongoing evolution in 5G called massive beamforming.  The practical deployment of 4G and 5G cellular networks with extensive system evaluation and field results have shown the superiority of beam-space multiplexing under some engineering limitations, such as implementation complexity and deployment scenarios. Lastly, we attempted to present four potential research directions of beam-space multiplexing in 6G and beyond.}

\vspace{0.4cm}
\section{acknowledgment}

The authors would like to give thanks to Prof. Dake Liu of Beijing Institute of Technology and the anonymous reviewers for reviewing the manuscript.


\begin{thebibliography}{99}

		\bibitem{chen2015comprehensive}{}
     S. Chen \emph{et al.,} {``A Comprehensive Survey of TDD-based Mobile Com-
	munication Systems from TD-SCDMA 3G to TD-LTE (A) 4G and 5G Directions,''}
 \emph{China Commun}., vol. 12, no. 2,  pp. 40-60, Feb. 2015.

	\bibitem{R1-091513}{}
R1-091513, {``System Evaluation on Dual Layer Beamforming,''} {CATT}, 3GPP TSG RAN WG1 meeting \#56bis, March 2009.



	\bibitem{TR36912}{}
3GPP TR 36.912 v10.0.0.0, {``Feasibility Study for Further Advancements  
	for E-UTRA (LTE-Advanced) (Release 10)
	,''} March 2011. 


\bibitem{ChenADPT}{}
S. Chen \emph{et al.,} {``Adaptive Beamforming in TDD-based Mobile Com-
	munication Systems: State of the Art and 5G Research Directions,
	''} \emph{IEEE Wireless Commun}., vol. 23, no. 6,  pp. 81-8, Dec. 2016.  



	\bibitem{GTI}{}
GTI White Paper, {``TD-LTE: Features and Performance,''} Version  4.3-1,  
Jul. 2011.
 
 \bibitem{Tom}{}
T. L. Marzetta, {``Noncooperative Cellular Wireless with Unlimited  
	Numbers of Base Station Antennas,
 	''} \emph{IEEE Trans.Wireless Commun}., vol. 9, no. 11, pp. 3590-3600, Nov. 2010.  
 
 

 
\bibitem{Chen2014}{}
S. Chen \emph{et al.,} {``The Requirements, Challenges, and Technologies for 5G 
	of Terrestrial Mobile Telecommunication,''} \emph{IEEE Commun. Mag}., vol.   
52, no. 5, pp. 36-43, May 2014.
  
\bibitem{Tse}{}
 D. Tse  \emph{et al.}, {Fundamentals of Wireless Communication}, Cambridge  
 University Press, 2005.
 
 \bibitem{Yangperformace}{}
H. Yang and T. L. Marzetta, {``Compressed Sensing-aided Downlink Channel Training    
	for FDD Massive MIMO Systems,''}  \emph{IEEE J.  
	Sel. Areas Commun}., vol. 31, no. 2, pp. 172-179, Feb. 2013.


\bibitem{HanComp}{}
Y. Han \emph{et al.,} {``Performance of Conjugate and   
		Zero-forcing Beamforming in Large-scale Antenna Systems,''}  \emph{IEEE Trans. Commun}., vol. 65, 
no. 7, pp. 2852-2862, Jul. 2017.




\bibitem{R1-164256}{}
3GPP R1-164256, {``NR MIMO based on Channel Reciprocity,''} CATT,           
3GPP TSG RAN WG1 meeting \#85bis, May 2016.


\bibitem{RR1-166478}{}
3GPP R1-166478, {``MIMO Transmission Schemes for NR,''} CATT,     
3GPP TSG RAN WG1 meeting \#86bis, Aug. 2016.


\bibitem{Emil}{}
	E. Bj{\"o}rnson \emph{et al.,} {``Massive MIMO is a Reality-What is Next? Five 
		Promising Research Directions for Antenna Arrays,''}  \emph{arXiv:1902.07678v1}, Feb. 2019. 
{

\bibitem{KBL}{}
K. B. Letaief \emph{et al.,} {``The Roadmap to 6G-AI Empowered Wireless Networks,''}  \emph{IEEE Commun. Mag.}, vol. 57, no. 8, pp. 84-90, Aug. 2019.  





\bibitem{Baoleiwang}{}
B. Wang \emph{et al.,} {``Spatial-and Frequency-Wideband Effects in 
	Millimeter-Wave Massive MIMO Systems,''}  \emph{IEEE Trans. Signal 
	Process.}, vol. 66, no. 13,  pp. 3393-3406, May 2018.

%
}






\end{thebibliography}
\end{document}